\documentclass[12pt,a4paper,american,fleqn]{article}
\usepackage{times}
\usepackage[T1]{fontenc}
\usepackage[latin1]{inputenc}
\usepackage{babel}
\usepackage[dvips]{graphics}
\usepackage{setspace}
\begin{document}

{\centering \textbf{The calculation of a normal force between multiparticle
contacts using fractional operators}\par}

{\centering J.S. Leszczynski\par}

{\centering \emph{Institute of Mathematics \& Computer Science, Technical
University of Czestochowa, 42-200~Czestochowa, ul.~Dabrowskiego~73,
Poland}\par}


\textbf{Abstract}

This paper deals with the complex problem of how to simulate multiparticle
contacts. The collision process is responsible for the transfer and
dissipation of energy in granular media. A~novel model of the interaction
force between particles has been proposed and tested. Such model allows
us to simulate multiparticle collisions and granular cohesion dynamics.

\textbf{1. Introduction}

Behaviours of granular materials have generated much interest in a~lot
of industrial processes involving transport of particles and also
in natural phenomena. The key aspect in such media is how to model
the interactions that may eventually take place between the particles.
The collision processes are responsible for the transfer and dissipation
of energy in granular materials. Moreover, the understanding of interaction
process is important in order to develop simulations and theoretical
studies. Discrete models much better simulate the collision process
than continuum models. We will focus on the molecular dynamics models
in which takes into account an expression for the repulsive force
acting between the particles. Particularly, we will analyse what happens
with the multiparticle contacts. Multiparticle interactions occur
when a~particle contacts with surrounding particles. Typical molecular
dynamics models~\cite{Cundall,Lee} are valid for particle collisions
being independent from one another. Nevertheless, we observe opposite
situation in behaviour of dense granular media especially for cohesive
particles. In literature~\cite{Pourin} notices the lack of energy
dissipation in molecular dynamics models. We will propose novel form
of a~force between contacting particles and we will investigate its
properties.

\textbf{2. Modelling multiparticle interactions}

We consider a~set of spherical particles moving under optional extortion.
We neglect a~particle rotation. Figure~1a shows situation when particles
move without any interactions. We describe a~distinguished particle
through its characteristic radius \( r_{i} \), mass \( m_{i} \),
position \( \mathbf{x}_{i} \) and linear speed \( \mathbf{\dot{x}}_{i} \).
Index \( i \) denotes a~given particle and varies from \( 1 \) to
\( np \), where \( np \) is a~total number of particles. Typical equation
of motion for one particle without collision can be written as
\begin{equation} \label{eq01}
 m_{i}\mathbf{\ddot{x}}_{i}=\sum ^{nc}_{l=1}\mathbf{F}_{l},
\end{equation}
where \( \mathbf{F}_{l} \) is an optional force and \( nc \) notices
a~total number of the optional forces. We can distinguish the force
as: gravitational one, drag force, etc. Figure~1b presents another
situation, this means multiparticle collisions. We add to Eq~(\ref{eq01})
a~sum of collisional forces and we have
\begin{equation} \label{eq02}
 m_{i}\mathbf{\ddot{x}}_{i}+\sum ^{nd_{i}}_{j=1\, \wedge j\neq i}
 \mathbf{P}_{j}=\sum ^{nc}_{l=1}\mathbf{F}_{l},
\end{equation}
where \( \mathbf{P}_{j} \) is a~collisional force acting between
two particles. Temporary index \( j \) denotes a~particle contacting
with the given particle \( i \) and \( nd_{i} \) is a~total number
of particles surrounding and contacting with the particle \( i \).
In the molecular dynamics models, particles virtually overlap when
a~contact occurs. Let \( C_{j} \) be a~point in which two particles
contact as shown on the detail \( A \) in Figure~1b. Let \( \Pi  \)
be a~plane tangent to colliding particles at the point \( C_{j} \)
and \( \zeta _{j} \) be a~normal direction to \( \Pi  \). We decompose
the collisional force into
\( \mathbf{P}_{j}=\mathbf{P}_{\xi \, j}+\mathbf{P}_{\eta \, j}+
 \mathbf{P}_{\zeta \, j} \),
where \( \mathbf{P}_{\zeta \, j} \) is the force acting on the normal
direction and \( \mathbf{P}_{\xi \, j} \), \( \mathbf{P}_{\eta \, j} \)
are forces acting on the tangent direction. In our considerations
we neglect the tangent forces assuming
\( \mathbf{P}_{\xi \, j}=\mathbf{P}_{\eta \, j}=0 \).
Here we present some examples of the normal force models which are
frequently applied ones in practical simulations. Cundall and
Strack~\cite{Cundall} proposed the force being a~linear combination of
viscous and elastic terms
\begin{equation} \label{eq03}
 \mathbf{P}_{\zeta \, j}=c_{j}\cdot \mbox{\boldmath{$\dot{\zeta}$}}_{j}+k_{j}
 \cdot \mbox{\boldmath{$\zeta$}}_{j},
\end{equation}
where \( k_{j} \) is a~stiffness of a~spring whose elongation is
\mbox{\boldmath{\( \zeta _{j} \)}} and \( c_{j} \) is a~damping
constant. According to the detail \( A \) presented in Figure~1b
we define the virtual overlap as
\begin{equation} \label{eq04}
 \mbox{\boldmath{$\zeta$}}_{j}=\left( r_{j}+r_{i}\right) \cdot
 \mathbf{e}_{\zeta \, j}-\left( \mathbf{x}_{j}-\mathbf{x}_{i}\right),
\end{equation}
and a~unitary vector \( \mathbf{e}_{\zeta \, j} \) normal to \( \Pi  \)
\begin{equation} \label{eq05}
 \mathbf{e}_{\zeta \, j}=\frac{\mathbf{x}_{j}-\mathbf{x}_{i}}
 {\left\Vert \mathbf{x}_{j}-\mathbf{x}_{i}\right\Vert }.
\end{equation}
Lee~\cite{Lee} investigated a~nonlinear version of
Eq~(\ref{eq03})
\begin{equation} \label{eq06}
 \mathbf{P}_{\zeta \, j}=c_{j}\cdot
 \mbox{\boldmath{$\dot{\zeta}$}}_{j}+k_{j}\cdot
 \left| \mbox{\boldmath{$\zeta$}}_j \right|^{\frac{3}{2}}\cdot
 \mbox {sign}\left( \mbox{\boldmath{$\zeta$}} _{j}\right).
\end{equation}
As a~remark, we want to comment that Eqs~(\ref{eq03}) and (\ref{eq06})
are typical for binary collisions. Several coefficients \( c_{j} \)
and \( k_{j} \) assumed as a~function of normal restitution coefficient.
In the case of multiparticle contacts we have to assume the same collisional
time which is not suitable for granular cohesion dynamics. Independently
on physical properties of granular materials, we cannot change surface
properties of contacting particles in above models. Therefore, we
cannot perform simulations taking into account the particles cohesion.
The restitution coefficient informs us about some work of deformation
between contacting bodies but does not inform how much time is needed
during particle collisions. In real behaviour of moving particles
we can easy change surface properties of granular materials. Especially,
when we consider the interactions in a~granular material for dry particles
and for wet ones. In crucial point of our discussion, we assume that
momentum and energy transfer between multiparticle contacts is identified
by \emph{memory effect}. From the other hand, a~given particle have
to remember about surrounding particles during collision process.
Using fractional calculus~\cite{Oldham} and the generalised viscoelastic
model~\cite{Schiessel} we propose a~novel form of the normal force
\begin{equation} \label{eq09}
 \mathbf{P}_{\zeta \, j}=c^{\alpha _{j}}_{j}\cdot
 k^{1-\alpha _{j}}_{j}\cdot \, _{t^{*}_{j}}D^{\alpha _{j}}_{t}
 \left( \mbox{\boldmath{$\zeta$}} _{j}\right),
\end{equation}
where \( c_{j} \) and \( k_{j} \) have the same meaning like in
previous models, \( \alpha _{j} \) is a~real order of differentiation
which belongs to the range
\( \alpha _{j}\in \left\langle 0\ldots 1\right\rangle  \) and
\( _{t^{*}_{j}}D^{\alpha _{j}}_{t}\left( \mbox{\boldmath{$\zeta$}} _{j}\right)  \)
is a~differential operator of the fractional order \( \alpha _{j} \).
According to~\cite{Oldham} we introduce a~definition of such operator
as left side Riemann-Liouville fractional derivative
\begin{equation} \label{eq10}
 _{t^{*}_{j}}D_{t}^{\alpha _{j}}\left( \mbox{\boldmath{$\zeta$}} _{j}\right) =
 \left\{ \begin{array}{lll}\frac{1}{\Gamma \left( n-\alpha _{j}\right) }
 \frac{d^{n}}{d\, t^{n}}\int\limits^{t}_{t^{*}_{j}}
 \frac{\mbox{\boldmath{$\zeta$}} _{j}
 \left( \tau \right) }{\left( t-\tau \right) ^{\alpha _{j}+1-n}}d\tau  &
 \mbox {for} & n-1<\alpha _{j}<n\\
 \frac{d^{n}}{d\left( t-t^{*}_{j}\right) ^{n}}
 \mbox{\boldmath{$\zeta$}} _{j}\left( t\right)  &
 \mbox {for} & \alpha _{j}=n
 \end{array}\right.,
\end{equation}
where \( n=\left[ \alpha _{j}\right] +1 \) and \( \left[ \cdot \right]  \)
denotes integer part of a~real number. We also have a~symbol \( t^{*}_{j} \)
which determines begin of the collision process. Taking into consideration
a~fact, given by Hilfer~\cite{Hilfer}, that the Riemann-Liouville
derivative has no physical interpretation - especially when we try
to introduce initial conditions - we formulate another one fractional
derivative
\begin{equation} \label{eq11}
 _{t^{*}_{j}}^{C}D_{t}^{\alpha _{j}}\left( \mbox{\boldmath{$\zeta$}}_{j}\right)
 =\left\{ \begin{array}{lll}\frac{1}{\Gamma \left( n-\alpha _{j}\right) }
 \int\limits^{t}_{t^{*}_{j}}
 \frac{\frac{d^{n}\mbox{\boldmath{$\zeta$}} _{j}
 \left( \tau \right) }{d\, \tau^{n} }}{\left( t-\tau \right)^{\alpha _{j}+1-n}}
 d\tau  & \mbox {for} & n-1<\alpha _{j}<n \\
 \frac{d^{n}}{d\left( t-t^{*}_{j}\right) ^{n}}
 \mbox{\boldmath{$\zeta$}} _{j}\left( t\right)  &
 \mbox {for} & \alpha _{j}=n
 \end{array}\right.,
\end{equation}
which is called Caputo derivative~\cite{Hilfer,Oldham}. According
to~\cite{Oldham} we describe transition between Eq~(\ref{eq10})
and Eq~(\ref{eq11}) in form
\begin{equation} \label{eq12}
  _{t^{*}_{j}}D^{\alpha _{j}}_{t}\left( \mbox{\boldmath{$\zeta$}} _{j}\right) =
 \sum ^{n-1}_{l=0}
 \frac{\left( t-t^{*}_{j}\right) ^{l-\alpha _{j}}}{\Gamma 
 \left( l-\alpha _{j}+1\right) }\cdot
 \mbox{\boldmath{$\zeta$}} _{j}\left( t^{*}_{j}\right) +\, 
 _{t^{*}_{j}}^{C}D^{\alpha _{j}}_{t}
 \left( \mbox{\boldmath{$\zeta$}}_{j}\right),
\end{equation}
where the sum means initial conditions. We explain with details a~meaning
and dependence between these two derivatives because many investigators
have concentrated on the Riemann-Liouville derivative, but this is
a~lack in physical applications. Everywhere in the contact process,
we numerically solve Eq~(\ref{eq02}) with formula~(\ref{eq09}) together
with Eq~(\ref{eq12}) by application the decomposition method described
in~\cite{Leszczynski}.

\textbf{3. Computational results}

To illustrate profits of the force model given by formula~(\ref{eq09})
we compare its behaviour with the force models presented by Eqs~(\ref{eq03})
and~(\ref{eq06}). In this case we simulate a~particle vertically
falling down to a~bottom plate as shown in Figure~2. When performing
this simulation, the particle falls under gravity (we used Eq~(\ref{eq01}))
and the contact occurs in the plate (we used Eq~(\ref{eq02})). Figure~2
presents vertical displacement of the particle over time. When the
collision occurs, we apply different schemes of the normal forces:
continuous line represents the linear force~(\ref{eq03}), dashed
line shows the nonlinear one~(\ref{eq06}) and two dash-dot lines
are responsible for our formula~(\ref{eq09}). For determined physical
properties of the particle and initial conditions we observe very
good agreement between our force model and the linear one. In such
case we establish the parameter \( \alpha  \) in Eq~(\ref{eq09})
\( \alpha =0.051 \). In the next case we can see quite good agreement
between our force model and the nonlinear one when \( \alpha =0.19 \).
We can simulate both cases changing the non-integer order \( \alpha  \)
in the fractional operator. In analysis of such behaviour we can observe
that the parameter \( \alpha  \) models the surface properties. When
\( \alpha =0 \) we have an~elastic collision but \( \alpha =1 \)
determines a~viscous collision. We notice that the material parameters
like \( c_{j} \) and \( k_{j} \) in formula~(\ref{eq09}) are constant.
Therefore, the fractional order \( \alpha  \) is a~surface parameter
balancing between elastic and viscous properties of two contacting
bodies. Moreover, formula~(\ref{eq09}) is suitable for multiparticle
contacts. Figure~3 presents some behaviour of three particles in two
dimensional space when the parameter \( \alpha  \) changed from \( 0.01 \)
to \( 0.9 \). Such simulation is so far from reality because we neglect
gravity force, tangential forces under particle contacts and particle
rotations. Thin lines represents particle trajectories but thick lines
are common trajectories when particles move like a~one body. In the
low values of \( \alpha  \) we do not observe common trajectories.
When the parameter \( \alpha  \) increases up to \( \alpha =0.9 \)
we can see common trajectories of two or three particles in dependence
on mutual positions. We strongly notice that formula~(\ref{eq09})
is more flexible to model cohesion processes than other ones. Extending
our considerations we observe different lengths of the common trajectories.
This fact issues from different begins of contacting times \( t^{*}_{j} \)
between two interacting particles. Therefore we can simulate multiparticle
contacts in which we do not assume the same collisional time.

\textbf{4. Conclusions }

We have proposed and discussed a~novel model of the normal force used
in simulations of the particle collisions. With this model, it is
possible to simulate multiparticle contacts in which we do not assume
the same collisional time. This feature, in comparison to the linear
and nonlinear force models is advantage of some generalisation between
particle interactions. However, some of the parameters of this model
may still be tuned, as for example, the parameter \( \alpha  \) in
order to reflect real surface properties of contacting particles.

\def\refname

\newpage
{\centering \resizebox*{1\textwidth}{!}{\includegraphics{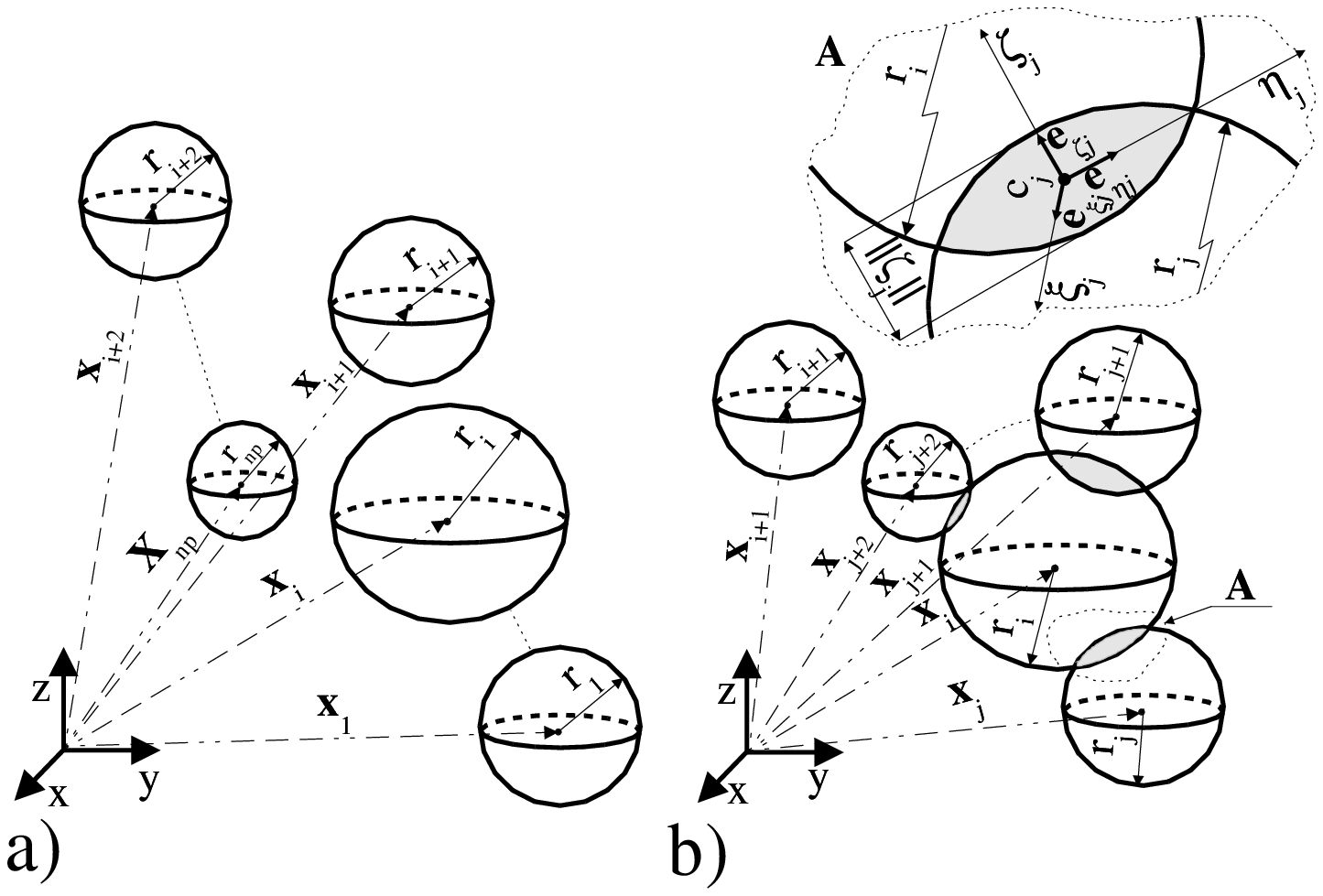}} \par}

{\centering Figure 1\par}

{\centering Sketch to illustrate particles behaviour: a) without collision;
b) with multiparticle contacts.\par}

\newpage
{\centering \resizebox*{1\textwidth}{!}{\includegraphics{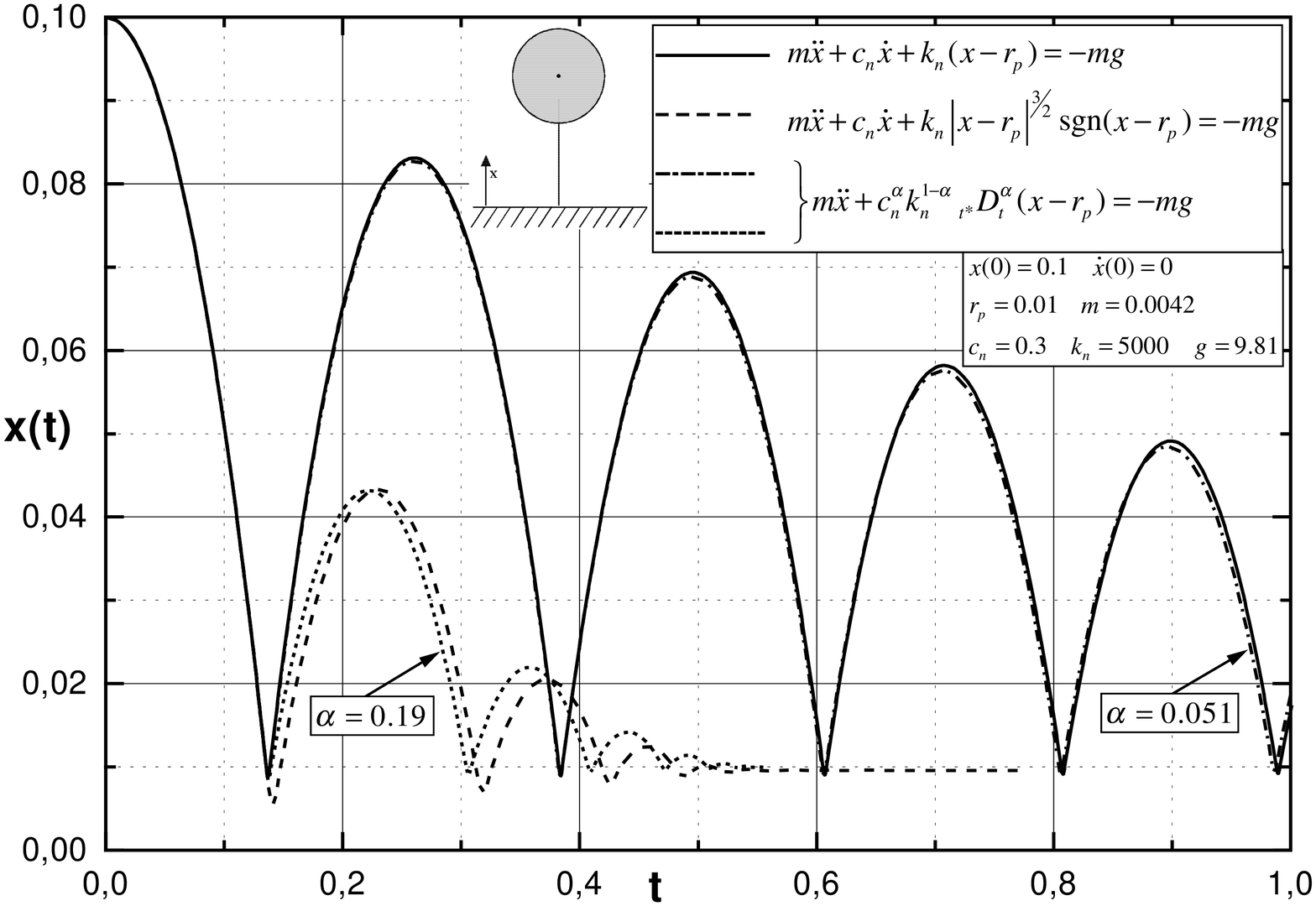}} \par}

{\centering Figure 2\par}

{\centering Displacement over time of a particle falling down on a
bottom plate.\par}

\newpage
{\centering \resizebox*{1\textwidth}{!}{\includegraphics{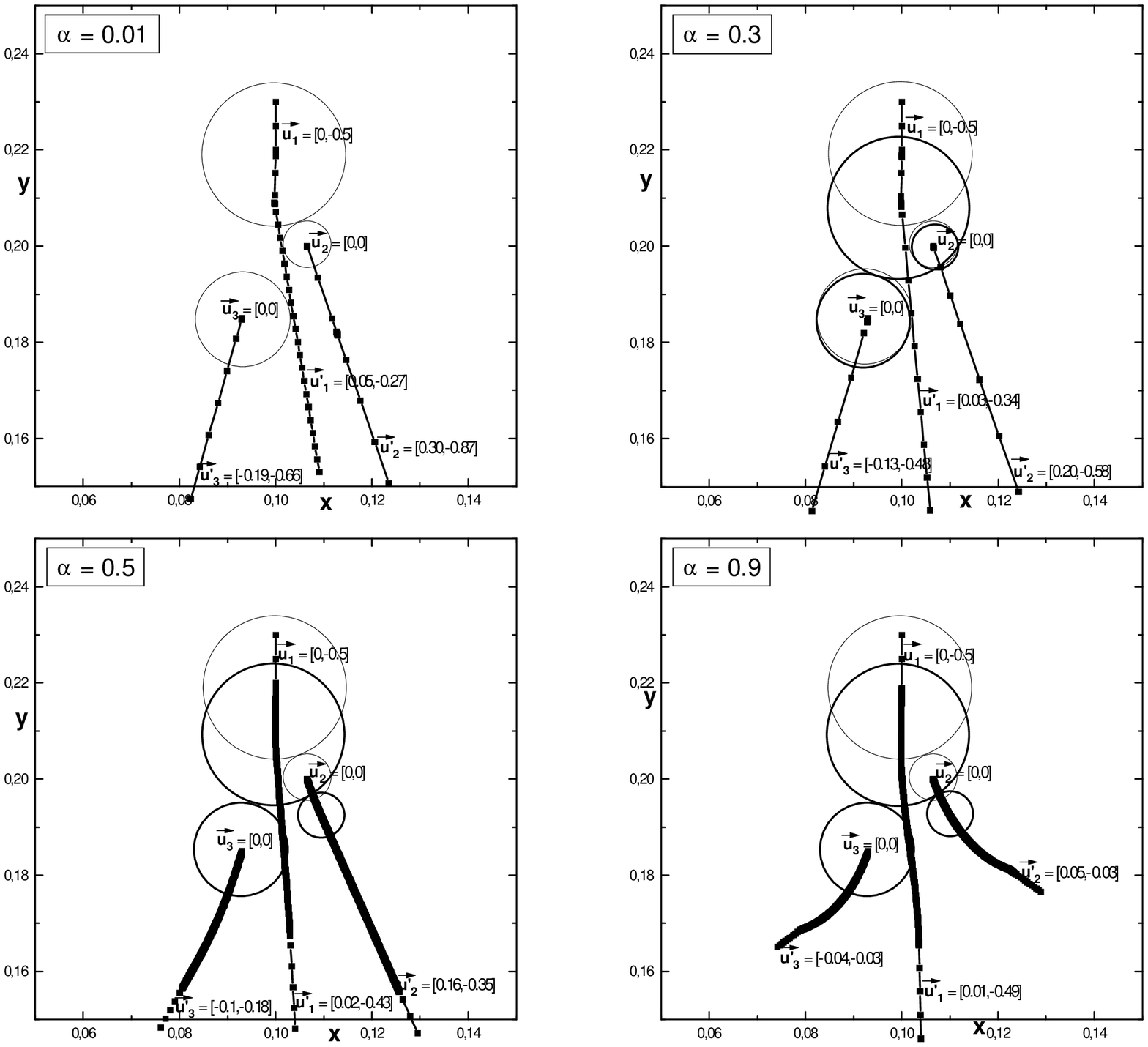}} \par}

{\centering Figure 3\par}

\centering Multiparticle contacts in dependence on surface properties.

\end{document}